\documentclass[journal=jacsat,manuscript=article]{achemso}

\usepackage{chemformula} 
\usepackage[T1]{fontenc} 
\usepackage{ amssymb }

\author{Carla S. Perez-Martinez}
\author{Susan Perkin}
\email{susan.perkin@chem.ox.ac.uk}
\affiliation[Oxford University]
{Physical and Theoretical Chemistry Laboratory, Department of Chemistry, University of Oxford, Oxford, OX1 3QZ}

\title[An \textsf{achemso} demo]
  {Interfacial structure and boundary lubrication of a dicationic ionic liquid}

\abbreviations{IR,NMR,UV}

\begin{document}

%
%
%
%
%

\begin{abstract}
We report measurements of the normal surface forces and friction forces between two mica surfaces separated by a nano-film of dicationic ionic liquid using a Surface Force Balance. The dicationic ionic liquid 1,10-bis(3-methylimidazolium)decane di-[bis(trifluoromethylsulfonyl)]imide forms a layered structure in nano-confinement, revealed by oscillatory structural forces. Friction measurements performed at different film thicknesses display \textit{quantized friction}, i.e. discontinuities in friction as layers are squeezed out and friction coefficients dependent on the number of liquid layers confined between the surfaces. The details of the friction traces indicate a liquid-like film, and, surprisingly, decreasing friction with increasing water content; we discuss possible mechanisms underlying these observations. This latter trait may be helpful in applications where ionic liquid lubricants cannot be insulated against humid environments.
The article is dedicated to the memory of Jacob N. Israelachvili.
\end{abstract}

\section{Introduction}
The intricate nano-structuring of ionic liquids in the bulk, at interfaces, and in confinement has been revealed through many experimental and theoretical investigations over the past decade or so\cite{Lopes:2006uw,Russina:2012cn,Fedorov:2014er,Hayes2015,smith2013c}. Composed entirely of charged species, ionic liquids exhibit structure determined by Coulombic constraints as well as van der Waals, steric and specific (e.g. hydrogen-bonding) interactions. Directly related to their nano-scale ordering and structure, ionic liquids display unusual dynamic properties such as distinct dynamic environments for charged and neutral solutes\cite{Araque:2015di}. \\
In confined environments, such as thin films, the nano-structure and dynamics of ionic liquids are closely entwined. The structure in confined films of ionic liquids depends strongly on the nature of the walls: if the confining surfaces are polar or charged, the ionic liquid typically stratifies into layers with (excess) counter charge opposing the surface charge, and oscillating charge density through subsequent layers decaying in magnitude away from the surface\cite{Zhou:2012fx,smith2013c,Fedorov:2014er,Hoth2014}. The dynamic properties of these layered films are interesting;  this has been investigated both in terms of the drainage of the film during compression between approaching surfaces\cite{Garcia2018,Lhermerout2018}, and in terms of the friction or lubrication properties between shearing surfaces\cite{Lhermerout2018a}. Experiments investigating the mechanical properties of nano-confined ionic liquids revealed finite yield stress and negative slip length\cite{Smith2013,Lhermerout2018} -- characteristic of a solid film -- yet with low friction coefficients, not dissimilar to 2D solid lubricants. A particular feature of sheared ionic liquid films, of significance for tuning dynamic surface interactions, is \textit{quantized friction}: distinct friction coefficients characteristic of each layer number and discontinuities in friction magnitude during compression\cite{Smith2013,Smith2014,Lhermerout2019b}. \\
A mechanistic understanding of lubrication and thin-film properties of ionic liquids is important for many of their applications. Ionic liquids can be designed for individual tasks by selection and combination of cation and anions to tune the properties; most generally, ionic liquids exhibit good thermal and (electro)chemical stability, negligible vapor pressure, capacity to resist high loads, moderate viscosity and conductivity. These favorable traits are leading to incorporation of ionic liquids in many technological directions, and here we focus on their increasing use as lubricants.  In particular, it has been demonstrated that dicationic ionic liquids (with monovalent anion, i.e. valence ratio 2:1) are particularly favourable for tribological applications because of their high decomposition temperature \cite{Zhang2017,Maton2013} and good anti-wear performance, especially at higher temperatures \cite{Jin2006,Zeng2008,Yu2008,Palacio2009,Mahrova2015,Nevshupa2017,Yu2007}. While several studies have measured the lubrication response of 1:1 ionic liquids in the boundary regime \cite{Lhermerout2018a,Smith2013,Smith2014,Espinosa-Marzal2014,Werzer2012,Elbourne2013}, the friction response of nanoconfined ionic liquids with ions of higher valencies remains largely unexplored\cite{Palacio2009}. The structure of nano-scale films of 2:1 ionic liquids might be expected to differ from that of 1:1 ionic liquids, and this should -- in turn -- lead to variation in the dynamic properties. \\
Here, we report our measurements carried out with a Surface Force Balance (SFB) to explore the interfacial arrangement, electrostatic interactions, and the boundary friction response of the dicationic ionic liquid 1,10-bis(3-methylimidazolium)decane di[bis(trifluoromethylsulfonyl)]imide, valence ratio 2:1, shortened as [C$_{10}$(C$_1$Im)$_2$][NTf$_2$]$_2$ (Fig.~1(a)).  We find that the dicationic ionic liquid forms distinct layered structures in confined films, as for 1:1 ionic liquids, with layer dimensions consistent with an arrangement whereby the cation lies with its long axis in the plane of the surfaces.  The film shows traits of a liquid, in particular inability to sustain a shear stress over $\sim$ second timescales. Friction forces measured across the film are quantized, friction coefficients are low, and -- most strikingly -- decrease when the film incorporates water from a humid environment. This latter trait is likely to be beneficial for dicationic ionic liquid lubricants in applications where complete insulation from environmental humidity is not possible.   

\begin{figure} 
\includegraphics{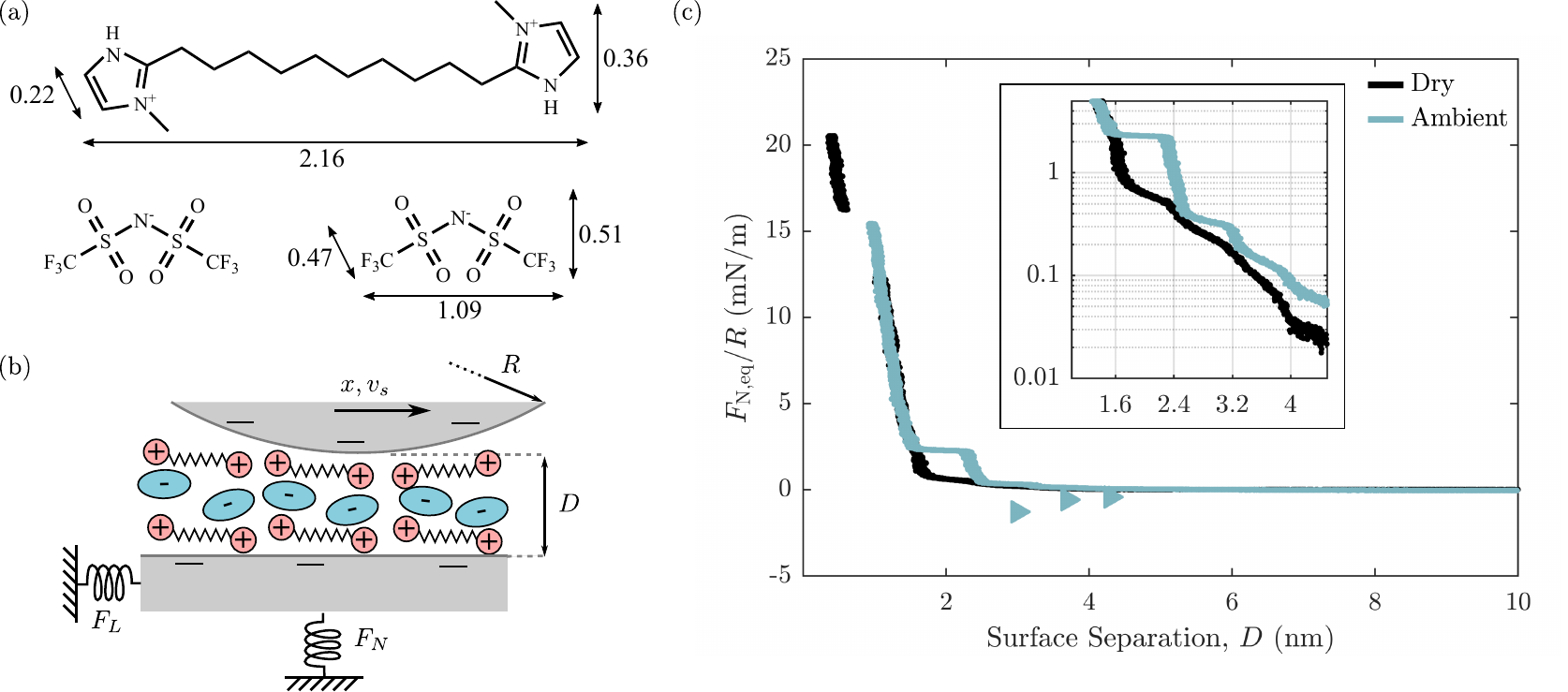}\caption{\label{Fig:ForceProfiles}  (a) 1,10 decane bis(3-methylimidazolium)decane di-bis(trifluoromethanesulfonyl)imide, [C$_{10}$(C$_1$Im)$_2$][NTf$_2$]$_2$. The ion dimensions are indicated in nanometers. The cation size is 2.16$\times$0.22$\times$0.36~nm (stretched), whereas the anion is 1.09$\times$0.47$\times$0.51~nm. (b) Schematic of the surface force balance (SFB) setup (c) Force profiles for dry and ambient equilibrated ionic liquid. We plot the equilibrium contribution to the force, i.e. $F_{N\rm{eq}} = F_N -F_{N,h}$. The hydrodynamic force contribution, $F_{N,h}$ was subtracted from the total measured force $F_N$ as described in the text and elaborated in the supplementary information.  The inset shows the detail of the force at small distances. For the outermost layers (located approximately at $D=2.4, 3.2$ and 4~nm), the adhesive minima are plotted as triangles in Figure~\ref{Fig:ForceProfiles}(c). Strong adhesion was detected when retracting the surfaces from the innermost layers (at about $D=1.6$ and 0.8~nm), but could not be measured exactly. Retraction data from the innermost layer and a small discussion are included in the supplementary information. The data shown are from a single measurement, however the layering and trends discussed in the text were reproduced in three separate experiments (separate mica sheets and liquid samples) and at different contact spots on the mica sheets. }
\end{figure}

\section{Experimental Section}
The SFB is used to measure the interaction forces between two surfaces separated by the ionic liquid (Fig.~\ref{Fig:ForceProfiles}(b)). The setup is explained in detail elsewhere, \cite{Klein1998} and recent modifications to our experimental setup are described in ref.~{\cite{Lhermerout2018}. In brief, two back-silvered mica sheets of equal thickness (ca. 3 $\mu$m) are glued onto cylindrical lenses and mounted in crossed-cylinder configuration with a radius of curvature $R~\sim$~1~cm. A droplet of ionic liquid is injected in between the atomically smooth, step-free mica surfaces. White light interferometry is used to determine the thickness $D$ of the ionic liquid film with a precision of 0.1 nm. One of the lenses is mounted on a piezo-electric tube which can be used to provide both normal displacement as well as lateral (sliding) motion. The lenses are mounted on a set of orthogonal leaf springs, a horizontal spring for measuring normal force ($k_N$= 125$\pm$6.4~N/m), and a vertical spring for measuring shear forces ($k_S$=441$\pm$4~N/m). By measuring the deflection of these springs, it is possible to determine the normal interaction force $F_N$ and the lateral or shear force $F_L$ as a function of $D$. 

The ionic liquid (Covalent Associates), is dried in vacuo ($<5\cdot10^{-3}$~mbar) at 80~$^\circ$C for at least twelve hours prior to injection. 
P$_2$O$_5$ is used as a desiccant inside the SFB chamber to keep the ionic liquid as dry as possible during the measurements. Once dried measurements have been performed, the chamber is allowed to equilibrate with ambient humidity. This equilibration process takes several hours, and so we take measurements during the equilibration process to evaluate the changes of normal forces and lubrication behaviour with increasing water content.

Measuring the equilibrium interaction forces for dicationic ionic liquids is challenging due to the high viscosity of these substances (0.7206~Pa$\cdot$s for [C$_{10}$(C$_1$Im)$_2$][NTf$_2$]$_2$\cite{Shirota2011}). This large viscosity induces a significant hydrodynamic force when performing continuous approaches between the surfaces. Chan and Horn \cite{Chan1985} first developed a method for modeling these viscous effects in simple molecular liquids in the SFB experimental geometry. The hydrodynamic force can be modelled using Reynolds' theory of lubrication, taking into account a negative slip boundary condition of the range of the structural force. Lhermerout and Perkin \cite{Lhermerout2018} recently showed a similar approach is also valid for treating ionic liquids, and they used this method to disentangle hydrodynamic and electrostatic forces in SFB measurements of monocationic ionic liquids of moderate viscosity ($\sim$~0.03~Pa$\cdot$s). We have used the methodology of Ref. \citenum{Lhermerout2018} to calculate the hydrodynamic forces in these experiments; details are provided in the supplemental information. We henceforth define the hydrodynamic force $F_{N,h}$, and the equilibrium interaction force between the surfaces $F_{N,\rm{eq}}$. The total normal load between the surfaces, assuming no electrokinetic effects (i.e., equilibrium electrostatic and hydrodynamic contributions are not coupled), is $F_N=F_{N,h}+F_{N,\textrm{eq}}$. 

Friction measurements are performed at several shearing velocities $v_s$  (470, 105 and 32~nm/s). Shearing experiments result in a measurement of the lateral force between the surfaces, $F_L$, as a function of time $t$ during the shearing cycle (of duration $t_{\rm{cycle}}$). We define the kinetic friction force, $F_S$, as the value of the lateral force when the surfaces reach continuous sliding. Note that we report all friction measurements against $F_N$, and not against $F_{N,\rm{eq}}$, since $F_N$ is the actual load acting between the surfaces. Friction measurements were performed using both stepped and continuous approaches. In the stepped case, the surface separation $D$ was reduced in discrete steps, and at each $D$, shear measurements were performed at different velocities for the same value of the total normal load $F_N$. In the continuous approach, the surfaces are brought together at a constant normal velocity (0.1-0.2~nm/s) while shearing the film continuously at $v_s=470~$nm/s\cite{Lhermerout2019b}. The shearing cycle is much faster than the change in normal distance, so that, during each single shearing cycle, the normal force and surface separation are effectively constant. The normal force profiles measured using a continuous approach with shearing are the same as when approaching the surfaces continuously without shearing, confirming there is no shear-induced layering of the ionic liquid. 

\section{Results and Discussion}
We begin by inspecting the measured equilibrium interaction force, $F_{N,\rm{eq}}$, between the surfaces across the dicationic ionic liquid as a function of their separation distance, $D$, as shown in Figure~\ref{Fig:ForceProfiles}(c).  
At large distances the surfaces experience no interaction above our experimental resolution until they reach separations $D < 5 ~\rm{nm}$. Then, at separations closer than 5~nm, the surfaces experience an alternating sequence of repulsive and attractive interactions which increase in magnitude as the separation decreases. Due to the mechanical spring-deflection method for detecting the force, leading to a spring instability when $dF_N/dD\geq k_N$, these interactions are observed as a series of repulsive `walls' separated by discontinuous jumps on the approach of the surfaces, with the attractive wells and minima measured during retraction of the surfaces. The damped oscillatory equilibrium interaction profile is commonly termed a \textit{structural force}, and its origins in the squeezing-out of sequential layers of molecules is well established for non-polar liquids\cite{Horn1981}, electrolytes \cite{MartinJimenez1ff}, and 1:1 ionic liquids\cite{Horn1988fd,smith2013c}. The magnitude, decay and wavelength of the oscillatory structural force gives insight into the strength of the liquid-solid interaction and the molecular arrangement in the film\cite{coles2018}.   \\
Here, we find that an oscillatory structural force is also present across the  2:1 ionic liquid [C$_{10}$(C$_1$Im)$_2$][NTf$_2$]$_2$ between two mica surfaces, qualitatively similar to the oscillatory structural forces observed previously with 1:1 ionic liquids. Two features of the $F_{N,\rm{eq}}$ vs. $D$ profile are particularly notable: (i) the wavelength of oscillations and its implication for structure in the film, and (ii) the strong effect of absorbed water on the magnitude of oscillations. We now consider each of these in turn. \\
 
The wavelength of oscillations is found to be 0.8$\pm$0.1~nm, which is very similar to that previously reported for imidazolium-based 1:1 ionic liquids with the same [NTf$_2$] ion and short alkyl-chains: For [C$_n$C$_1$(Im)][NTf$_2$] with $n = 2$ or 4, wavelengths or repeat-distances for the structural force have been reported in the range 0.71--0.8~nm \cite{Cheng2016,Han2018,Lhermerout2018,smith2013c}. These values are similar to the dimension of an ion pair (or, equivalently, to the ++ or - - correlation length in the bulk fluid), and so have been interpreted as the repeat-distance for a layered film consisting of (on average) alternating layers of cations and anions which are expelled as ion-pair-layers when the surfaces approach\cite{smith2013c}. As long as the alkyl chain is sufficiently small, the value of $n$ was found not to alter the wavelength, indicating that the alkyl chain is oriented in the plane of the film\cite{Smith2013a}.  In contrast to this consistent wavelength for small $n$, it was found that 1:1 ionic liquids with longer alkyl chains give rise to much larger wavelength -- \mbox{typically $\gtrsim$ 2~nm} -- indicating larger repeat-distances in the layered film; this was interpreted as bilayers with segregated alkyl-chain-regions and ionic regions\cite{smith2013c,Cheng2016,Griffin2017,Freitas2018}.  
It is perhaps surprising, then, that the dication studied here, with its $n=10$ alkyl chain, retains a wavelength of 0.8~nm; indistinguishable from the wavelength of short-chain imidazolium 1:1 ionic liquids. This strongly suggests that the dication is oriented with its long axis in the plane of the surfaces, and therefore indistinguishable in height compared to monovalent analogues such as [C$_4$C$_1$(Im)][NTf$_2$]; the two charged sites on each cation lie in the same plane of the layered structure, rather than `straddling' two layers or any other perpendicular orientation. A diagrammatic representation of the proposed structure with cations lying in the plane of the layers is in Figure~\ref{Fig:ForceProfiles}(b). Related to this finding, it has been noted in the past that the presence of two charged sites on an ionic liquid cation results in less pronounced nano-scale segregation compared to monovalent ionic liquids, and thus it might be expected that the dicationic ionic liquid will be less prone to form bilayers in nano-confinement\cite{Hayes2015}.

A second feature of the normal force measurements shown in Figure~\ref{Fig:ForceProfiles}(c) is the strong effect of water on the magnitude of surface forces. The forces required to squeeze out layers of liquid in the highly confined film are substantially greater for the ambient-equilibrated liquid compared to the dry ionic liquid (see inset to Figure~\ref{Fig:ForceProfiles}(c)). The amplification of the oscillatory structural force also leads to detection of surface interactions extending to larger separation distances; in the example shown in Figure~\ref{Fig:ForceProfiles}(c) three additional minima were recorded after the film was allowed to absorb water by equilibration with ambient atmosphere compared to the measurement with dry liquid. Similar behaviour has been reported in the past with 1:1 ionic liquids\cite{Cheng2016}. The strong enhancement of ionic liquid nano-structure due to small fractions of water (often at a level undetectable by analytical methods such as Karl-Fischer titration) is well known. In this case of enhancement of the force between mica surfaces, there are two possible contributions: (i) increase of the bulk nano-structure due to water (solvophobic self-assembly), which is detected by way of the structural forces, and (ii) increase of the mica surface charge in the presence of water, which leads to increase of the surface-charge-induced structuring (overscreening). The first of these mechanisms would be expected to modify the surface forces between confining walls of any material, whereas the latter is relevant only for surfaces which charge-regulate by adsorption/desorption of ions (as is the case with mica). \\ 

A final point to consider in inspecting the normal surface forces relates to the recent observations of long-range surface forces in ionic liquids, beyond the oscillatory forces, called \textit{underscreening}\cite{Gebbie:2015fn,Smith2016,lee2017FD}. In the past, all measurements of underscreening involved monovalent electrolytes, either 1:1 ionic liquids or 1:1 salts in polar solvent. It is therefore of interest to ask the question as to whether underscreening is observed in the 2:1 ionic liquid (and how the apparent decay length scales with ion valence). However, in our present measurements with [C$_{10}$(C$_1$Im)$_2$][NTf$_2$]$_2$, we are not able to distinguish any force above the experimental noise level at distances beyond the oscillatory region.  There are several possible interpretations of this: first, the underscreening could be of shorter range in 2:1 than in 1:1 ionic liquids, so that the force is obscured by the structural forces which are 1-2 orders greater in magnitude. Secondly, the effective surface charge at the outside of the oscillatory (overscreening) region could be closer to zero, leading to unmeasurably low amplitude of electrostatic forces (despite finite decay length). Thirdly, the high viscosity of these 2:1 ionic liquids, which necessitate careful subtraction of dynamic viscous forces from the measured interaction (as described above and in the SI), may obscure a small additional electrostatic interaction. The question of electrostatic forces in highly concentrated electrolytes at higher valence remains pertinent and could be addressed in future experiments with 2:1 salts in polar solvent of lower viscosity. \\

\begin{figure}
\includegraphics{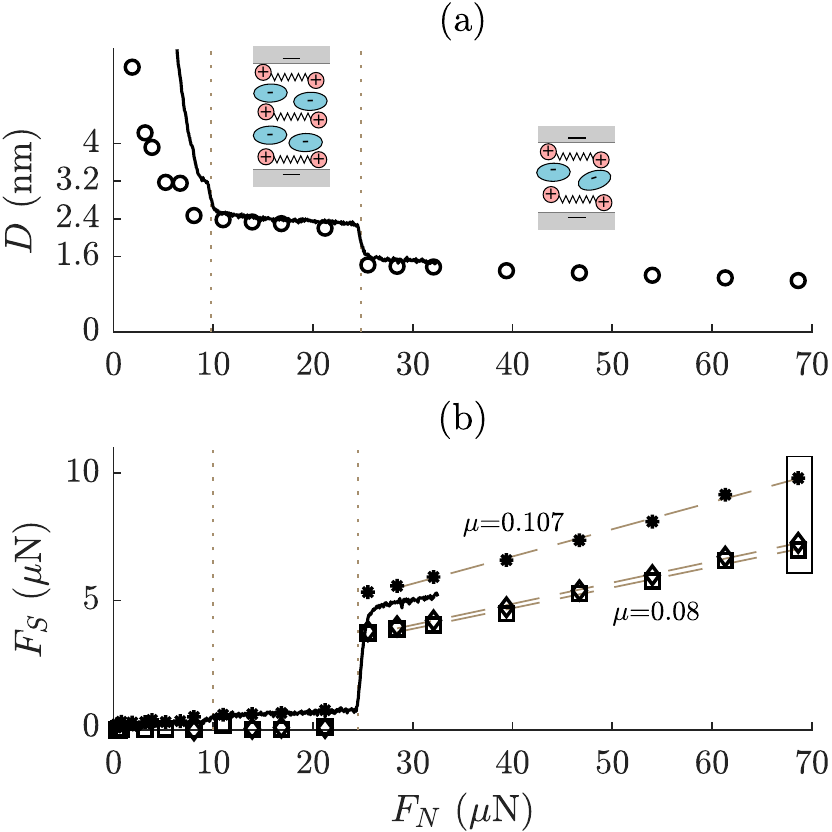}\caption{\label{Fig:FrictionOverview} (a) Surface separation $D$ as a function of normal force $F_N$ during friction measurements for the ambient equilibrated ionic liquid. The continuous line corresponds to an approach performed while shearing continuously at $v_S=470~$nm/s, whereas the circles represent the distances at which shear measurements where performed during a stepped approach. (b) Kinetic friction force $F_S$ as a function of $F_N$, for a continuous approach while shearing at $v_S=470~$nm/s (solid line), and for measurements in the stepped approach (markers). The squares, diamonds and asterisks represent measurements at $v_s=$~470,~105,~and 52 nm/s, respectively. The points within the rectangular box at 70~$\mu$N are those obtained from the raw traces shown in Figure~\ref{Fig:Frictiondetail}.}
\end{figure}

Next, we address the friction forces measured across the thin ionic liquid films, which we measure as a function of varying load, liquid film thickness, and incorporated water.  Figure~\ref{Fig:FrictionOverview} shows the separation distance, $D$, (in (a)), and the kinetic friction force, $F_S$, (in (b)), each as a function of normal force, $F_N$. It is clearly seen that steps in $D$ arising from squeeze-out of molecular layers as the normal force $F_N$ is increased are accompanied by simultaneous steps in $F_S$. These steps in $F_S$ are characterised by discontinuities in both the magnitude of $F_S$ and in the gradient, or friction coefficient $\mu = dF_S/dF_N$. The series of discrete quasi-linear friction-load regions, each corresponding to a different number of ion layers (or film thicknesses), has been called \textit{quantized friction} as reported in the past for 1:1 ionic liquids\cite{Smith2013a,Werzer2012,Elbourne2013}. \\
For film thicknesses $D>$3.2~nm, the measured friction is extremely low and below our instrumental resolution for all $v_s$. When the normal force is increased sufficiently to expel a further molecular layer - decreasing the surface separation to $D=2.4$~nm - the friction force becomes measurable and increases approximately linearly with normal force (at sufficiently large shearing velocity; see below). Then, when the normal force is increased further, another molecular layer is squeezed out, resulting in surface separation of $\sim D=1.6$~nm, for which the value of $F_S$ is larger and increases (again quasi-linearly) with a steeper gradient.\\
 
The magnitude and gradient of $F_S$ is found to be mildly dependent on velocity. At a separation of ca. 2.4 nm, the friction is still within the instrument noise for $v_s$ of 52 and 105~nm/s, but a friction response can be detected at $v_S=470~$nm/s. On the next layer, ca. 1.6 nm, $F_S$ is of the order of a few $\mu$N for all velocities probed, and the friction coefficient shows a very weak velocity dependence ranging from $\mu = 0.08$ to $\mu = 0.11$ with one order of magnitude increase in shearing velocity. We note that our two different methods for detecting the shear force - continuous approach whilst shearing and stepped approach - give rise to friction measurements in good agreement (as compared by the solid lines and markers in Figure~\ref{Fig:FrictionOverview}).\\
 
\begin{figure}
\includegraphics{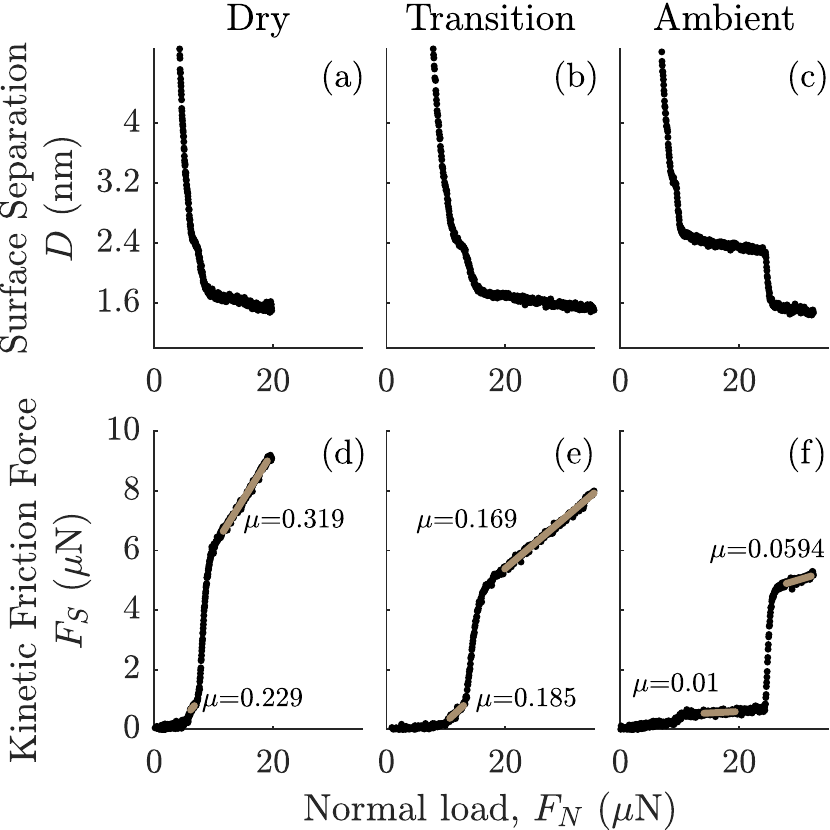}\caption{\label{Fig:FrictionWetvsDry} (a)-(c) Surface separation $D$ and (d)-(f) kinetic friction force $F_S$ as a function of total normal load $F_N$, for the dry, transition and ambient-equilibrated ionic liquid.}
\end{figure}

We find that the friction measured across the molecularly thin dicationic ionic liquid films is strongly affected by the incorporation of small quantities of water from the ambient atmosphere. Figure~\ref{Fig:FrictionWetvsDry} shows the concomitantly-measured surface separation and friction force as a function of applied normal load for three continuous approaches carried out at different points during the process of absorbing water from the atmosphere. The approaches were performed at normal velocities of 0.1-0.2~nm/s, and the shearing velocity was 470 nm/s for all cases shown. It is clear, from the data for the layer at ca. 1.6~nm, that (i) the friction coefficient fitted from the data \emph{decreases} with increasing water content, and that (ii) a greater value of normal load is required to reach the thinnest film with increasing water content. This latter point has the important corollary that \emph{friction reduction can be achieved by resistance to squeeze out}, for systems where a thicker film has lower friction coefficient (as is always observed with IL lubricants here and in the past, but is not the case for simple alkane films). In the present experiment, higher humidity leads to a higher barrier for squeeze-out of IL layers, and therefore the friction, and friction coefficient,  can be more than an order of magnitude lower than for the dry liquid at the same value of applied load (e.g. compare $F_S$ and $\mu$ at 20~$\mu$N for dry vs. ambient liquid). \\

\begin{figure}
\includegraphics{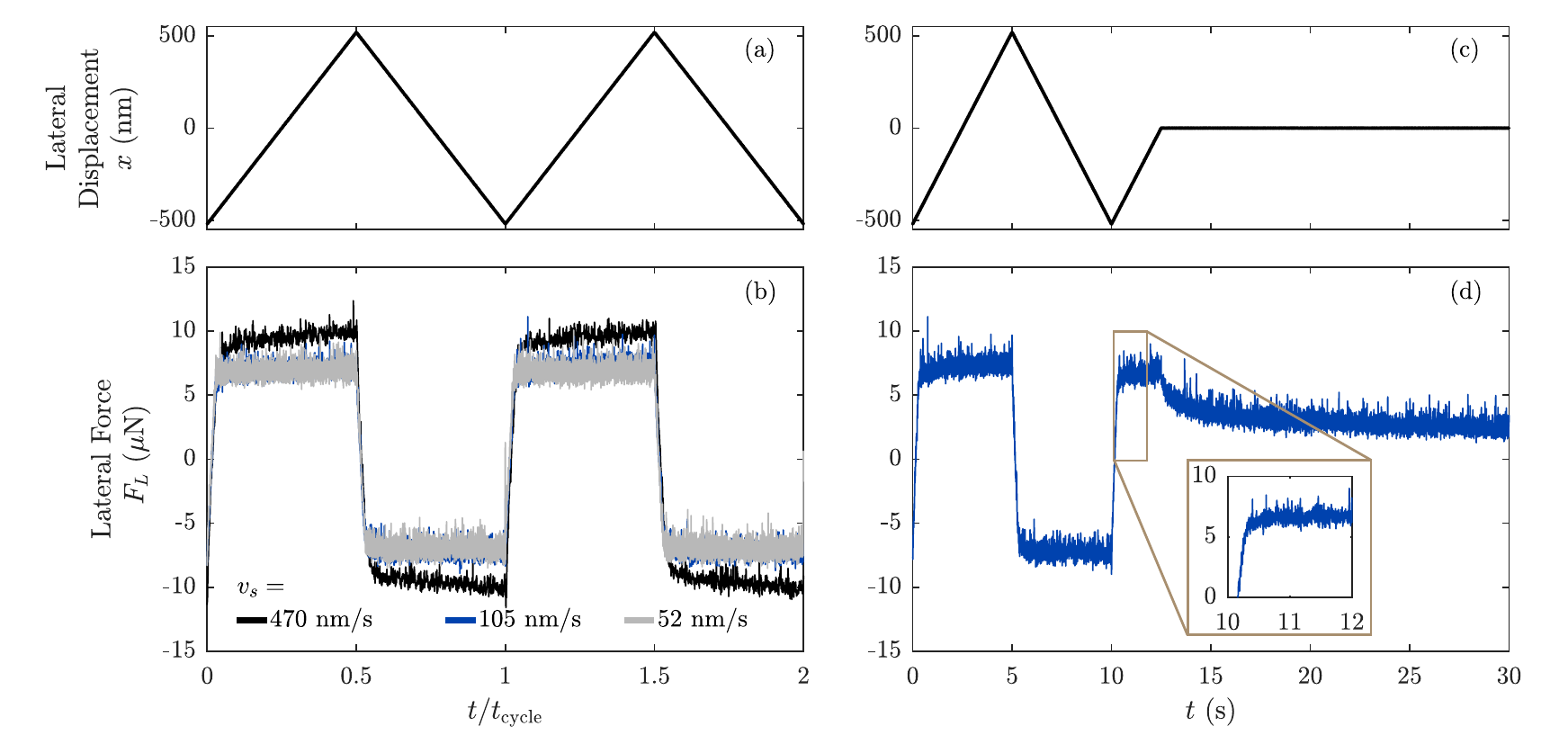}\caption{\label{Fig:Frictiondetail} (a)-(b) Friction response as a function of shearing velocity and (c)-(d) example of stress relaxation. (a) Applied lateral displacement~$x$ and (b)~lateral force~$F_L$ at three different shearing velocities. Data is shown for two shearing cycles. For all three traces shown, $D=$1.1~nm and $F_N=68.7~\mu$N. These data correspond to the highlighted data points in Figure~\ref{Fig:FrictionOverview}(b).  (c) Applied lateral displacement~$x$ and (d)~lateral force~$F_L$ as a function of time during switching off of lateral displacement. The lateral force shows a relaxation of the film after shearing is stopped. For this curve, $v_S=$52~nm/s, $D=$1.1~nm and $F_N=68.7~\mu$N. }
\end{figure}

Next, to obtain even closer insight into the behaviour of the dicationic ionic liquid film during shearing, we inspect details of the friction traces recorded during individual shearing cycles. An example of the lateral force, $F_L$ measured as a function of time during two shearing cycles is shown in Figure~\ref{Fig:Frictiondetail}(a),(b), and the shear force relaxation after halting the constant-velocity displacement is shown in Figure~\ref{Fig:Frictiondetail}(c),(d).  In these measurements the lateral displacement, $x$, is varied linearly with time (constant velocity) in a back-and-forth motion (Figure~\ref{Fig:Frictiondetail}(a)). The resulting $F_L$ trace shows that stress is built up across the film until the (kinetic) shear force of the film, $F_S$, is reached; at which point the surfaces slide past each other. We observe no measurable yield stress at the initiation of sliding, and there is no measurable stick-slip behaviour. Furthermore, on closer inspection of the transition from coupling to sliding - as seen in the expanded region inset to Figure~\ref{Fig:Frictiondetail}(d) - we see that there is curvature in $F_L$ \textit{vs.} $t$ rather than a discontinuity (yield point), indicating finite shearing of the film below $F_S$. This is an indication that the thin (1.1~nm) film, which we suppose to consist of 3 layers of ions (cations-anions-cations, as illustrated in Figure~\ref{Fig:ForceProfiles}(c)), is not entirely solid-like, and instead has features of a soft or liquid-like film. \\
This conclusion that the film is liquid-like is firmly supported by the relaxation behaviour observed when halting the shear motion, as in  Figure~\ref{Fig:Frictiondetail}(c),(d). At the point of switching off lateral shear motion (at 12.5~s) there is finite $F_L$, then, over the course of ca. 10 seconds, this stress is relaxed and $F_L$ returns towards zero. This slow but clear relaxation indicates a \textit{viscous} state for this 2:1 ionic liquid. This contrasts with earlier observations with some 1:1 ionic liquids which exhibited clear yield stress, stick-slip and solid-like characteristics\cite{Smith2013}, and shows that the precise viscoelastic properties of these nano-scale films are sensitively dependent on the molecular details.   \\
To finish, we consider an interpretation of these two key observations: (i) the striking enhancement of lubrication effect (reduction of friction) with the incorporation of ambient water, and (ii) the detailed insights that the ionic liquid film remains as a viscous liquid - rather than solid - under confinement.  The observation that water \textit{reduces} the friction coefficient in [C$_{10}$(C$_1$Im)$_2$][NTf$_2$]$_2$ is particularly notable for its contrast with several previous cases of boundary lubrication with 1:1 ionic liquids, where water tended to increase friction \cite{Lhermerout2018a}. Espinosa-Marsal showed that the friction increases for the ionic liquids [C$_2$C$_1$Im][EtSO$_4$], [C$_2$C$_1$Im][FAP] and [C$_6$C$_1$Im][FAP]  when the liquids were saturated with ambient humidity \cite{Espinosa-Marzal2014}, and experiments with two bilayers of [C$_{10}$C$_1$Pyrr][NTF$_2$] showed an increase of the friction coefficient by more than an order of magnitude when water was present within the layers \cite{Smith2014}. A recent colloid probe study on [C$_{2}$C$_1$Im][NTf$_2$] reported an increase of friction for the liquid equilibrated with 44\% RH compared to the dry liquid \cite{Han2018}. Molecular dynamics simulations also showed an increase in friction with added water \cite{Fajardo2017}.  In those cases, the films were often determined to be solid-like; i.e. with some elastic characteristics; shearing the film is then an activated process, either involving translation of layers across one another or shear-melt. The addition of water to the film is well understood to enhance the structure of an ionic liquid, as discussed already above in relation to the normal forces. For solid films, the enhancement of structure is clearly expected to also increase friction, since the activation barriers increase; sliding involves motion across a more strongly corrugated energy landscape. For liquid-like films, the addition of water still enhances the nano-structure (c.f. the strength of layering increases Fig.1c), but now this appears to be connected with a reduction in friction. This may be due to more well-defined ion layers in the $z$-direction - which creates low-stress planes for shearing, perhaps involving collections of fluorinated anions in particular  - without creating in-plane structure. This rationale is drawn out in schematic form in Figure~\ref{Fig:Frictionexplanation}. Alternatively (or additionally), it may be that the water molecules act to screen the direct electrostatic interaction between ions and thereby diminish the Coulombic resistance to shearing. This latter mechanism is similar to the reduction in viscosity of bulk ILs by the addition of water.\\
\begin{figure}
\includegraphics{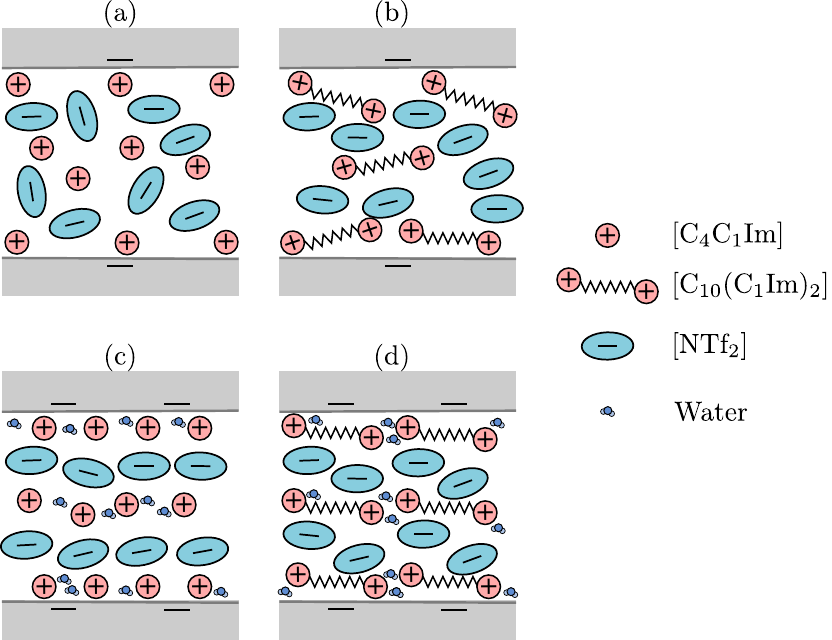}\caption{\label{Fig:Frictionexplanation} Hypothesis as to the effect of water on surface charge and structuring in the monocationic and dicationic ionic liquids between mica surfaces. (a) Dry 1:1 ionic liquid; (b) dry 2:1 ionic liquid; (c) hydrated 1:1 ionic liquid, (d) hydrated 2:1 ionic liquid. The effect of hydration on the mica surface charge has been reported in the literature and discussed in the text; the influence of this on the ionic liquid structure and how this differs for 2:1 ionic liquids is suggested in the text and illustrated here for clarity.} 
\end{figure}

\section{Summary and Conclusions}
In summary, we have presented a detailed study of the normal and shear forces acting between charged surfaces across nano-scale films of a dicationic ionic liquid. The liquid is found to form a layered structure in the film, with strongly adsorbed layers preventing squeeze-out of the liquid until substantially higher pressures than for equivalent un-charged, non-polar, liquids. The force profile shows the characteristic traits of a structural force, i.e. step-wise increases in force on approach of the surfaces and multiple energy minima on retraction. The shear force across the film was measured as a function of load, revealing quantized friction regimes with friction coefficients dependent on the number of ion layers between the surfaces.  Detailed analysis of the friction traces shows the film to have characteristics of a viscous material: finite shear below the limiting value of sliding friction, and spontaneous relaxation of stress.  Surprisingly, the friction and friction coefficient reduced when the dry ionic liquid was allowed to incorporate small fractions of water from the atmosphere. This latter trait is likely to be helpful in applications where low friction is desired and complete insulation against the atmosphere is not possible.

\begin{acknowledgement}

The authors thank Paulo Lozano and Catherine Miller (MIT) for providing us with samples of the ionic liquid. Romain Lhermerout is acknowledged for his important contributions to aspects of the methodology; details of which are to be reported separately. An anonymous referee is thanked for a helpful suggestion regarding the effect of water in screening ion interactions.  Funding is gratefully acknowledged from The Leverhulme Trust (RPG-2015-328) and the European Research Council (under Starting Grant No. 676861, LIQUISWITCH).

\end{acknowledgement}

\begin{suppinfo}

Details on the calculation of hydrodynamic forces, a discussion of ion bridging during retraction of the surfaces, and other experimental details are included in the supplementary information.


\end{suppinfo}

\bibliography{Dications_refs.bib}

\end{document}